%% file: SSthesis.tex
\documentclass[12pt]{report}
\usepackage{amssymb,amsfonts,amsmath}
\usepackage{graphicx}
\usepackage{indentfirst}
\usepackage{fancybox}
\usepackage{color}
\usepackage[square, comma, numbers, sort&compress]{natbib}

\parskip=\medskipamount

\def\z{\zeta}

\def\tr{{\rm tr}}

\newcommand{\be}{\begin{equation}}
\newcommand{\ee}{\end{equation}}
\newcommand{\bea}{\begin{eqnarray}}
\newcommand{\eea}{\end{eqnarray}}

\newcommand{\ket}[1]{\, |#1\rangle}
\newcommand{\bra}[1]{\langle #1 |\,}

\def\double #1{#1{\hbox{\kern-2pt $#1$}}}

\usepackage{chngcntr}

\counterwithin{figure}{chapter}
\counterwithin{equation}{chapter}
\counterwithin{table}{chapter}

\begin{document}

\bibliographystyle{unsrt}

\pagenumbering{roman}

\input{titlepage.tex}

\input{isbn.tex}

\input{abstract.tex}

\newpage
\hfill\\

\input{acknowledgements.tex}

\input{contents.tex}
\input{listofpub.tex}

\input{introduction.tex}

\input{chap01.tex}

\input{chap02.tex}

\input{chap03.tex}

\input{conclusions.tex}

\input{bibliography.tex}
\end{document}

%% file: titlepage.tex
\newpage
\thispagestyle{empty}

\begin{flushleft}
UNIVERSITY OF HELSINKI \hspace{\fill}  REPORT SERIES IN PHYSICS
\end{flushleft}

\begin{center}

\vspace{5mm}

\large  HU-P-D161

\vspace{28mm}

{\Large \bf Quantum Space-Time and Noncommutative \\
Gauge Field Theories}

\vspace{0.9cm}

{\Large  Sami Saxell}

\vspace{2.3cm}

{\normalsize
Division of Elementary Particle Physics \\
Department of Physics \\
Faculty of Science \\
University of Helsinki\\
Helsinki, Finland
}

\end{center}

%% file: isbn.tex
\newpage
\thispagestyle{empty}

\hfill\\
\vfill

%% file: abstract.tex
\newpage
\setcounter{page}{1}

\section*{Abstract}

\noindent Arguments arising from quantum mechanics and gravitation theory as well as from string theory, indicate that the description of space-time as a continuous manifold is not adequate at very short distances.
An important candidate for the description of space-time at such scales is provided by noncommutative space-time 
where the coordinates are promoted to noncommuting operators. Thus, the study of quantum field theory in noncommutative space-time provides an interesting interface where ordinary field theoretic tools can be used to study the properties of quantum space-time. 

\noindent The three original publications in this thesis encompass various aspects in the still developing area of noncommutative quantum field theory, ranging from fundamental concepts to model building.
One of the key features of noncommutative space-time is the apparent loss of Lorentz invariance that has been addressed in different ways in the literature. One recently developed approach is to eliminate the Lorentz violating effects by integrating over the parameter of noncommutativity. Fundamental properties of such theories are investigated in this thesis.
Another issue addressed is model building, which is difficult in the noncommutative setting due to severe restrictions on the possible gauge symmetries imposed by the noncommutativity of the space-time. Possible ways to relieve these restrictions are investigated and applied and a noncommutative version of the Minimal Supersymmetric Standard Model is presented.
While putting the results obtained in the three original publications into their proper context,
the introductory part of this thesis aims to provide an overview of the present situation in the field.

%% file: acknowledgements.tex
\newpage
\section*{Acknowledgements}

\noindent This thesis is based on research work done at the Division of Elementary Particle Physics, Department of Physics, University of Helsinki. I gratefully acknowledge the grant from GRASPANP, the Finnish Graduate School in Particle and Nuclear Physics. I acknowledge also the grants from the Magnus Ehrnrooth Foundation and from the University of Helsinki Mathematics and Science fund.

\noindent First of all, I want to thank my supervisors, Docent Anca Tureanu and Professor Masud Chaichian, for their invaluable guidance during my time as a Ph.D. student. Their
enthusiasm and positive attitude created an inspiring and enjoyable working atmosphere vital for good scientific research work.
I express my gratitude to the referees of this thesis, Associate Professor Archil Kobakhidze and Professor Dmitri Vassilevich, for careful reading of the manuscript and for useful comments.
I wish to thank my colleagues, most notably Dr.\ Masato Arai and Dr.\ Nobuhiro Uekusa, for fruitful collaboration, as well as invaluable support and encouragement.

\noindent I am grateful also to my friends and office-mates for discussions about physics and other topics.
Especially I would like to thank my colleagues and friends, Paavo and Tommi, for many discussions and long-lasting friendship.

\noindent Finally, I would like to express my deepest gratitude to my parents, my brother and my beloved wife, for their constant love and support.

\vspace{10mm}
\noindent Helsinki, April 2009

\vspace{5mm}
\noindent {\it Sami Saxell }

%% file: contents.tex
\tableofcontents

%% file: listofpub.tex
\chapter*{List of original publications}
\markboth{LIST OF ORIGINAL PUBLICATIONS}{LIST OF ORIGINAL PUBLICATIONS}
\addcontentsline{toc}{chapter}{\protect\numberline{}List of original publications}

\begin{enumerate}

\item[I.]  {\it ``A noncommutative version of the minimal supersymmetric standard model,''}\\
	M.~Arai, S.~Saxell and A.~Tureanu, \\
   Eur.\ Phys.\ J.\  C {\bf 51}, 217 (2007)
  [arXiv:hep-th/0609198]. 

\item[II.] {\it ``Circumventing the No-Go Theorem in Noncommutative Gauge Field Theory,''}\\
 M.~Arai, S.~Saxell, A.~Tureanu and N.~Uekusa, \\
   Phys.\ Lett.\  B {\bf 661}, 210 (2008)
  [arXiv:0710.3513 [hep-th]].

\item[III.] {\it ``On general properties of Lorentz invariant formulation of noncommutative quantum field theory,''}\\
S.~Saxell, \\
  Phys.\ Lett.\  B {\bf 666}, 486 (2008)
  [arXiv:0804.3341 [hep-th]].

\end{enumerate}

%% file: introduction.tex
\chapter{Introduction}
\pagenumbering{arabic} 
In the standard quantum field theory, space-time is treated as a continuum that can be described mathematically as a differentiable manifold. While this approach gives an adequate picture of space-time at large length scales, there is no evidence that this description should hold to arbitrarily small scales. 
In fact, the continuum description leads to divergences that plague ordinary quantum field theory. While physical quantities can be calculated despite the divergences using renormalization theory, ordinary quantum field theory cannot be considered as a complete description of fundamental physics. 
These divergences provide a signal that the continuum description of space-time should be replaced by some new structure at short distances of the order of Planck length
 $\lambda_p\approx 1.6\times 10^{-35}\,m$, where quantum gravitational effects, if nothing else, should modify the concept of space-time. The most prominent candidate for the UV completion of continuum quantum field theory is string theory. In string theory the point-particle description of quantum field theory is replaced by vibrating strings which provide an effective minimal length given by the string length. Another approach to the UV physics is to keep the quantum field description of particles but instead replace the space-time manifold  itself by some structure that exhibits a minimal length. This is the underlying idea in noncommutative quantum field theory (NC QFT) where the smeared nature of space-time is obtained by replacing the space-time coordinates by a noncommutative algebra.

Noncommutative algebra is an old concept in mathematics and it is familiar to physicists from its natural appearance in various physical contexts such as nonabelian symmetry groups and quantum mechanics. In quantum mechanics, the classical position and momentum variables are replaced by operators $\hat x^i$ and $\hat p_i$ in a Hilbert space, that satisfy the canonical commutator algebra $\left[\hat x^i , \hat p_j\right]=\delta^i_j$. Points in the classical configuration or momentum space correspond to eigenvalues of the operators $\hat x^i$ or $\hat p_i$, respectively,  but since the coordinate and momentum operators do not commute, they do not share the same eigenstates and thus the points in the classical phase-space get smeared out. The study of such quantum spaces was pioneered by von Neumann, whose studies led to the theory of von Neumann algebras and to the birth of "noncommutative geometry", in which the study of spaces is done purely in algebraic terms \cite{connesbook}. 

The generalization of the canonical commutators of quantum mechanics to nontrivial commutators between the coordinate operators was suggested by Heisenberg \cite{heisenberg}.
The first published paper on the subject \cite{snyder} is by H. S. Snyder who proposed an algebra of the form
\begin{eqnarray}
\left[\hat x^\mu , \hat x^\nu\right]=ia^2/\hbar L_{\mu\nu},
\label{}
\end{eqnarray}
where $a$ is a parameter that defines the basic unit of length and $L_{\mu\nu}$ are the generators of the Lorentz group.
Snyder's space-time is symmetric under Lorentz transformations but not under translations. Later, the formulation was modified  by C. N. Yang \cite{yang} in order to achieve full Poincar\'e invariance. The motivating idea in Snyder's work was to eliminate the ultraviolet divergences arising in quantum field theory by making the space-time pointless on small length scales. However, due to the success of the renormalization program that was being developed at around the same time, Snyder's approach did not become popular.

The mathematical study of noncommutative geometry was revived in the 1980's most notably by Connes, Woronowicz and Drinfel'd, who generalized the notion of differential structure to the noncommutative setting \cite{conn}. 
The development of differential calculus in noncommutative geometry gave rise to physical applications such as Yang-Mills theory on NC torus \cite{torus} and the Connes-Lott model \cite{conneslott} 
based on Kaluza-Klein mechanism where the extra dimensions are replaced by noncommutative structures. The aim of the Connes-Lott model is to obtain a geometrical interpretation for the fields and various parameters in the Standard Model.

More concrete motivation for space-time noncommutativity came more recently from the work of Doplicher, Fredenhagen and Roberts who combined the quantum mechanical
uncertainty principle between coordinates and momenta with the classical Einstein's gravity theory \cite{Doplicher:1994zv}. This results in a limit on the accuracy
by which space-time measurements can be performed and thus ordinary space-time loses operational meaning at very short distances. It turns out the uncertainty relations 
can be described by a non-vanishing commutator for the coordinates,
\begin{eqnarray}
\left[ \hat x^\mu, \hat x^\nu \right]=i\hat Q^{\mu\nu}.
\label{DFRcommutator}
\end{eqnarray}
The right hand side of (\ref{DFRcommutator}) is a tensor operator that commutes with the coordinates and leads to uncertainty relations between the coordinates similar to the Heisenberg uncertainty relations of quantum mechanics.

Another impetus to the study of noncommutative field theory came from string theory. As it was shown by Seiberg and Witten in \cite{Seiberg:1999vs}, if the endpoints
of open strings are confined to propagate on a $D$-brane in a constant $B$-field background, then the endpoints live effectively on noncommutative space whose coordinates satisfy
the commutation relations
\begin{eqnarray}
\left[ \hat x^\mu, \hat x^\nu \right]=i\theta^{\mu\nu},
\label{commutatorintro}
\end{eqnarray}
where $\theta$ is a constant matrix. The dynamics of the open strings in the low energy limit is then described by NC QFT.
In this approach  $\theta^{\mu\nu}$ is a constant parameter that leads to breaking of Lorentz-invariance in the NC space-time.

One of the characteristic properties in the study of noncommutative physics is the apparent lack of guiding principles. The vague hints coming from quantum gravity (string theory) and the classical gravity argument of Doplicher et al. leave a huge freedom when postulating the properties and principles of the NC space-time and field theory in it. 
For this reason, there is also a lot of controversy in fundamental issues such as: How is the NC space-time defined? Can the breaking of Lorentz invariance be allowed? Can the Lorentz-invariant causality condition be violated and if so, what should it be replaced with? These controversies are salient also in the works that comprise this thesis.
Although 
the present day formulations of field theory in NC space suffer from severe problems that make them  unlikely candidates for a realistic realization of the idea of NC space-time, they still offer valuable toy models due to the relatively simple changes that are required on ordinary quantum field theory tools. The field theories on NC space is the main object of study in this thesis.

The three publications presented in this thesis encompass various aspects of NC quantum field theory ranging from fundamental properties of NC space-time physics to gauge theories and phenomenological model building on the NC space-time. In the introductory part of this thesis we review some essential results in NC quantum field theory, at the same time placing the included papers in their proper context. In chapter 2 we review the basic formulation of field theories in NC space-time and discuss perturbative aspects of such theories.
In chapter 3 we review some essential properties of QFT in NC space-time including the issue of space-time symmetries, going also beyond the standard formulation of NC space-time that is based 
on the constant parameter of noncommutativity. Especially, we consider Lorentz invariant formulation of NC field theory and the problems that such theories posses as was shown in the paper {\bf III}. In chapter 4 we review some aspects of NC gauge theories and model building. Gauge theories in NC space-time obey a no-go theorem that restrains model building in this context. In the paper {\bf II} a way to circumvent the no-go theorem was considered.
 This widens the possibilities of NC model building and was used in the construction of NC MSSM in {\bf I}.

%% file: chap01.tex
\chapter{Quantum field theory on NC space-time}
In this chapter we will review the presentation of fields on NC space-time using Weyl symbols and the corresponding
Moyal *-product.
This technique was introduced by Weyl in ordinary quantum mechanics, providing a mapping from 
functions of the phase space to quantum operators \cite{Weyl}.  
In NC space-time this approach provides a convenient way to study field theories.
For the general theory of *-products and deformation quantization, see \cite{Kontsevich:1997vb} and references therein.
More recent developments can be found in \cite{Kupriyanov:2008dn, McCurdy:2008ew}.

\section{Fields in NC space-time and the *-product}

In this section we follow the exposition given in the review paper \cite{Szabo:2001kg}.
Consider the commutative algebra of complex-valued functions in Euclidean $\mathbb{R}^D$.
The functions are assumed to satisfy the Schwarz
condition
\begin{equation*}
\sup_x(1+|x|^2)^{k+n_1+...+n_D}|\partial_1^{n_1}...\partial_D^{n_D}|^2 f(x)<\infty
\hspace{2mm},\hspace{2mm}k,n_i\in\mathbb{Z}_+,\hspace{2mm}\partial_i=\partial/\partial
x^i,
\end{equation*}
which guarantees a rapid decrease at infinity. This condition allows for the functions
to be described by their Fourier transforms. The noncommutative
space is defined by replacing the coordinates $x^i$ by operators
$\hat{x}^i$ obeying the commutation relations 
\begin{eqnarray}
\left[ \hat x^\mu, \hat x^\nu \right]=i\theta^{\mu\nu}.
\label{commutator}
\end{eqnarray}
For now we are assuming that $\theta$ is an antisymmetric matrix of real constant elements, and refer to this case as the {\it canonical noncommutative space-time}.
The noncommutative coordinates $\hat x^\mu$ generate an algebra of noncommutative operators. Given a 
function with the Fourier transform
\begin{eqnarray}
\tilde{f}(k)=\int d^Dx \; e^{-ik_i x^i}\;f(x),
\label{}
\end{eqnarray}
we define the corresponding Weyl operator by
\begin{eqnarray}
\hat{W}[f]=\int \frac{d^Dk}{(2\pi)^D} \; \tilde f(k) e^{ik_\mu\hat{x}^\mu}.
\label{W(f)}
\end{eqnarray}
Here the exponential function is defined by its expansion with symmetric ordering. 
This correspondence provides a mapping between operators and fields and the field $f(x)$
is called the Weyl symbol of the operator $\hat W[f]$.
The mapping can be written explicitly using the Hermitian operator
\begin{eqnarray}
\hat\Delta(x)=\int \frac{d^Dk}{(2\pi)^D} \;  e^{ik_\mu\hat{x}^\mu} \; e^{-ik_\mu x^\mu}.
\label{}
\end{eqnarray}
Then
\begin{eqnarray}
\hat{W}[f]=\int d^Dx\;f(x)\hat{\Delta}(x).
\label{eka}
\end{eqnarray}

A set of derivatives can be introduced
through linear anti-Hermitian derivations satisfying
\begin{equation}
\left[\hat{\partial}_i,\hat
{x}^j\right]=\delta_i^j\;,\hspace{1cm}\left[\hat{\partial}_i,\hat{\partial}_j\right]=0.
\end{equation}
From this definition it follows upon integration by parts that
\begin{equation}
\left[\hat{\partial}_i,\hat{W}[f]\right]=\hat{W}[\partial_if],
\end{equation}
and also that translation generators can be represented by unitary operators
\begin{equation}
e^{v^i\hat{\partial}_i}\hat{\Delta}(x)e^{-iv^i\hat{\partial}_i}=\hat{\Delta}(x+v).
\end{equation}
This implies that any cyclic trace has the property that
Tr $\hat{\Delta}(x)$ is independent of $x$. Therefore the trace is
uniquely given by an integration over space-time,
\begin{equation}
\mbox{Tr } \hat{W}[f]=\int d^Dx \;f(x),
\end{equation}
where the trace has been normalized so that Tr $\hat{\Delta}(x)=1$. This implies that
the trace plays a role of integration over noncommuting
coordinates.
Using the Baker-Campbell-Hausdorff formula
\begin{equation}
 e^{ik_i\hat{x}^i}e^{ik'_i\hat{x}^i}=e^{\frac{i}{2}\theta^{ij}k_ik'_j}e^{i(k+k')_i\hat{x}^i},
\end{equation}
one obtains
\begin{eqnarray}
\hat{\Delta}(x)\hat{\Delta}(y)&=&\iint\frac{d^Dk}{(2\pi)^D}\frac{d^Dk'}{(2\pi
)^D}\;e^{i(k+k')_i\hat{x}^i}e^{-\frac{i}{2}\theta^{ij}k_ik'_j}e^{-ik_ix^i-ik'_iy^i} \label{DeltaDelta}\\
&=&\iint\frac{d^Dk}{(2\pi )^D}\frac{d^Dk'}{(2\pi )^D}\int
d^Dz\;e^{i(k+k')_iz^i}\hat{\Delta}(z)e^{-\frac{i}{2}\theta^{ij}k_ik'_j}e^{-ik_ix^i-ik'_iy^i}.
\nonumber
\end{eqnarray} 
Since $\theta$ is assumed to be an invertible matrix, the Gaussian integrations over $k$ and $k'$ can be performed. The result is
\begin{equation}
\hat{\Delta}(x)\hat{\Delta}(y)=\frac{1}{\pi^D|\mbox{det }\theta|}\int d^Dx\;\hat{\Delta}(z)\;e^{-2i(\theta^{-1})_{ij}(x-z)^i(y-z)^j}.
\label{toka}
\end{equation}
In particular, it follows using the trace normalization and antisymmetry of $\theta^{-1}$, that the operators $\hat{\Delta}(x)$ for $x\in\mathbb{R}^D$ form an orthonormal set,
\begin{equation}
\mbox{Tr }\left(\hat{\Delta}(x)\hat{\Delta}(y)\right)=\delta^D(x-y).
\label{kolmas}
\end{equation}
This implies that the transformation from $f(x)$ to $\hat W[f]$ is invertible with the inverse transform given by
\begin{equation}
f(x)=\mbox{Tr }\left(\hat{W}[f]\hat{\Delta}(x)\right).
\end{equation}
Therefore, the map $\hat{\Delta}(x)$ provides a one-to-one correspondence between fields and Weyl operators.

Using (\ref {W(f)}) and (\ref{DeltaDelta}) one can deduce that the mapping is an isomorphism if the product between the functions of $x^\mu$ is given by the Moyal *-product:
\begin{eqnarray}
f(x)* g(x)&=&\iint\frac{d^Dk}{(2\pi )^D}\frac{d^Dk'}{(2\pi )^D}\;\tilde{f}(k)\tilde{g}(k'-k)\;e^{-\frac{i}{2}\theta^{ij}k_ik'_j}e^{ik'_ix^i}\nonumber\\
&=&f(x)\;\mbox{exp }(\frac{i}{2}\overleftarrow{\partial_i}\,\theta^{ij}\,\overrightarrow{\partial_j})\;g(x)  \label{star}\\
&=&f(x)g(x)\hspace{-1mm}+\hspace{-1mm}\sum_{n=1}^\infty\left(\frac{i}{2}\right)^n\hspace{-1mm}\frac{i}{n!}\theta^{i_1j_1}\cdots\theta^{i_nj_n}\partial_{i_1}\cdots\partial_{i_n}f(x)\partial_{j_1}\cdots\partial_{j_n}g(x). \nonumber
\end{eqnarray}
An alternative representation for the *-product can be derived using (\ref{eka}), (\ref{toka}) and (\ref{kolmas}),
\begin{eqnarray}
f(x)*g(x)&=&\mbox{Tr } \left( \hat W [f] ~ \hat W [g] ~ \hat \Delta (x) \right) \nonumber \\
&=& \frac{1}{\pi^D |\mbox{det} \theta|} \int \; d^D y \; d^D z f(y)\; g(z) \; e^{-2i (\theta^{-1})_{ij}(x-y)^i(x-z)^j }.
\label{integralrep}
\end{eqnarray}
The *-product is associative but noncommutative, and it provides a realization of the noncommutative algebra in terms of functions on
ordinary space-time.
Especially, we note that the *-product realizes the commutator (\ref{commutator})
\begin{eqnarray}
[x^\mu , x^\nu]_* := x^\mu * x^\nu - x^\nu * x^\mu=i\theta^{\mu\nu}.
\label{}
\end{eqnarray}
Due to the correspondence between the operator trace and space-time integration the integral of the *-product,
\begin{eqnarray}
\int d^Dx ~ f_1(x)*\cdot \cdot \cdot * f_n(x) =\mbox{Tr }\left(\hat{W}[f_1]\cdot \cdot \cdot \hat{W}[f_n]\right),
\label{}
\end{eqnarray}
is invariant under cyclic permutation of the functions $f_i$.
We note also that in the case of two fields,
\begin{eqnarray}
\int d^Dx\;f(x)* g(x)=\int d^Dx\;f(x)\,g(x),
\label{intproperty}
\end{eqnarray}
which follows upon integration by parts.

The *-product imposes nonlocality into products of fields. If the fields $f$ and $g$ are supported over a small region of size $\delta\ll \sqrt{|| \theta ||}$, then $f*g$ is non-vanishing over a region
of size $||\theta||/ \delta $ \cite{Minwalla:1999px}. For example, two point sources described by Dirac delta functions get
infinitely spread due to the *-product:
\begin{eqnarray}
\delta^D(x)*\delta^D(x)=\frac{1}{\pi^D |\mbox {det} \theta|}.
\label{}
\end{eqnarray}
This nonlocality has significant consequences on perturbative field theory.


\section{Field theory and quantization}

Let us start by considering the noncommutative version of the $\phi^4$ scalar field theory. In this chapter we assume that time and space commute; subtleties arising from
nonvanishing commutator between time and space will be discussed in the next chapter.  
The action for the noncommutative fields is given by the trace of the noncommutative Lagrangian, 
\begin{eqnarray}
S=\mathrm{Tr}\left(\frac{1}{2}\left[\hat{\partial}_i,\hat{W}[\phi]\right]^2+\frac{m^2}{2}\hat{W}[\phi]^2+\frac{g}{4!}\hat{W}[\phi]^4\right).
\label{}
\end{eqnarray}
This action can be rewritten in terms of the Weyl symbols using the formulae of the previous section:
\begin{eqnarray}
 S=\int
    d^Dx\left[\frac{1}{2}\left(\partial_i\phi(x)\right)^2+\frac{m^2}{2}\phi(x)^2+\frac{g}{4!}\phi(x)*\phi(x)*\phi(x)*\phi(x)\right].
\label{}
\end{eqnarray} 

Note that due to the property (\ref{intproperty}), the part of the action corresponding to a free theory is equivalent to its commutative counterpart.
From here we can proceed with the quantization by the usual path integral procedure.
The bare propagator is unchanged, but the interaction term receives a noncommutative modification. To read off the expression for the interaction vertex, we
write the interaction term's Fourier expansion,
\begin{eqnarray}
\int d^Dx ~\phi_*^4 = \Pi _{a=1}^4\left( \int \frac{d^Dk_a}{(2\pi)^D}\tilde \phi (k_a)\right) (2\pi)^D \delta^D \left(\sum_{a=1}^4 k_a\right) V(k_1,k_2,k_3,k_4).\nonumber\\
\label{}
\end{eqnarray}
Here the interaction vertex is given by the phase factor
\begin{eqnarray}
V(k_1,k_2,k_3,k_4)=\Pi_{a<b} e^{-\frac{i}{2} k_a \wedge k_b} ~~\mbox{and}~~k_a \wedge k_b:=k_{a\mu}\theta^{\mu\nu}k_{b\nu}.
\label{}
\end{eqnarray}

One of the most peculiar new effects in noncommutative quantum field theory is the mixing of
ultraviolet and infrared degrees of freedom that arises in perturbative calculations \cite{Minwalla:1999px}.
This effect shows the nonlocality of NC field theories that is reminiscent of a similar property of strings. 
Let us consider the one loop mass renormalization in the $\phi_*^4$ theory. Taking $D=4$, the
Euclidean action reads
\begin{eqnarray}
 S=\int
    d^4x\left[\frac{1}{2}\left(\partial_i\phi(x)\right)^2+\frac{m^2}{2}\phi(x)^2+\frac{g}{4!}\phi(x)*\phi(x)*\phi(x)*\phi(x)\right].
\label{}
\end{eqnarray}
The one-loop correction to the 1-particle irreducible two point function can be conveniently divided
into the planar and nonplanar parts:
\begin{eqnarray}
\Gamma(p) &=& \Gamma_{\mbox{planar}}(p)+\Gamma_{\mbox{nonplanar}}(p) \nonumber \\
          &=& \frac{g}{3}\int \frac{d^4k}{(2\pi)^4}\frac{1}{k^2+m^2}+\frac{g}{6}\int\frac{d^4k}{(2\pi)^4}\frac{1}{k^2+m^2}e^{2ik\wedge
p}.
\label{}
\end{eqnarray}
The planar diagram is proportional
to the standard commutative contribution with UV divergence and can be computed
using the Schwinger parametrization:
\begin{eqnarray}
\Gamma_{\mbox{planar}}(p)&=\frac{g}{48\pi^2}\left( \Lambda^2-m^2 \ln (\frac{\Lambda^2}{m^2})+O(1)\right).
\label{}
\end{eqnarray}
The phase factor in the nonplanar diagram acts effectively as a cutoff, yielding,
\begin{eqnarray}
\Gamma_{\mbox{nonplanar}}(p)&=\frac{g}{96\pi^2}\left( \Lambda_p^2-m^2 \ln (\frac{\Lambda_p^2}{m^2})+O(1)\right),
\label{}
\end{eqnarray}
where 
\begin{equation*}
\Lambda_p=\frac{1}{1/\Lambda^2+p\circ p}   ~~~\text{and}~~~   p\circ p\equiv p_\mu p_\nu (\theta^2)^{\mu\nu}.
\end{equation*}
Note that the nonplanar part remains finite when $\Lambda\rightarrow \infty $.
However, the ultraviolet divergence is restored in the limit $p\circ p\rightarrow 0$ which is achieved  either by
taking the commutative limit or the IR limit $p\rightarrow 0$. 
By taking first $p$ to zero one recovers the standard mass renormalization,
\begin{equation}
m_\text{ren}^2=m^2+\frac{1}{32}\frac{g\Lambda^2}{\pi^2}-\frac{1}{32}\frac{gm^2}{\pi^2}\text{ln }\frac{\Lambda^2}{m^2}+O(g^2).
\end{equation}  
With nonzero $p\circ p$ the correction assumes a complicated form that cannot be attributed
to any mass renormalization. The UV ($\Lambda\rightarrow 0$) and IR ($p\rightarrow 0$) limits obviously do not commute, 
which shows a curious mixing between the low and high energy dynamics.
The pole in the nonplanar loop at $p=0$ comes from the high momentum region of integration as $\Lambda\rightarrow \infty$.

UV/IR mixing is a general NC effect that affects also NC gauge theories, where both quadratic and linear poles appear \cite{Matusis:2000jf}.
Supersymmetry cancels these poles at least at one loop level, but typically logarithmic divergences persist.
The UV/IR mixing is a perturbative effect that seems to spoil renormalizability. As a counterexample, a noncommutative scalar field theory that has the property of being renormalizable
was constructed in \cite{Grosse:2004yu}. In this model the renormalizability is obtained by introducing into the action a harmonic term that makes the theory
covariant under duality transformation between coordinates and momenta. The harmonic term obviously breaks translational invariance,
but progress towards a translational invariant formulation that preserves renormalizability and renormalizable NC gauge field theories has been made, see
\cite{Blaschke:2008jh} and references therein.

As the previous example in the noncommutative Euclidean space-time illustrates, noncommutativity itself
is not enough to remove UV-divergences in perturbation theory and the naive expectation that noncommutativity
might regularize the divergences is not fulfilled. 
However, since the short-distance effects are related to long-distance features, topological restrictions can change the convergence properties \cite{Chaichian:1998kp}.
In fact, while in the case of classical space-time the theories on a sphere or cylinder have UV-divergences, the theories in the fuzzy sphere \cite{hoppe, grosse} and quantum cylinder \cite{Chaichian:1998kp}
do not have divergences at all due to the compactness properties of the spaces. 
In any case, the mixing between long and short distance physics is
an interesting new property on its own behalf and it implies that the introduction of space-time noncommutativity
 indeed brings novel and exciting ingredients into quantum field theories.


\section{Approaches towards finite range NC}

The UV/IR mixing effect resembles the situation in ordinary quantum mechanics where large distance phenomena in half of the phase space coordinates (the momenta) are related to short distance phenomena in the other coordinates (spatial position).
The infrared divergence can be thought of as a signal of the nonlocality in space extending to infinite range.

 An improvement to this situation could thus be obtained even in a fully noncompact space if the nonlocality could be restricted to a finite range\footnote{The discussion presented in this section is based
on an ongoing work by the author with Anca Tureanu and Masud Chaichian.}.
It should be remembered that the Moyal product (\ref{star}) is not the only *-product realization of the noncommutative algebra and at least some nonlocal effects
could depend on the choice of the product. The Moyal product was obtained by working in the Weyl basis (\ref{W(f)}) with symmetric ordering. Choosing for example the coherent state basis, the *-product is given by the Wick-Voros product
\begin{eqnarray}
f*_{WV} g (x) = f e^{\theta \overleftarrow\partial_+\overrightarrow\partial_-} g (x).
\label{}
\end{eqnarray}
Here we have restricted the space-time to two dimension s for simplicity and denoted
\begin{eqnarray}
\partial _{\pm}= \frac{1}{\sqrt{2}}\left( \frac{\partial}{\partial x^1} \mp
\frac{\partial}{\partial x^2}\right).
\label{}
\end{eqnarray}
Calculating the free propagator for a scalar field theory defined with the Wick-Voros product seems to lead to an apparent improvement in the UV-behaviour of the theory. However, it turns out that  the improvement does not persist when interaction amplitudes are calculated \cite{Chaichian:1998kp}. For a more recent study of the Wick-Voros product in NC QFT, see \cite{Galluccio:2008wk}.

A naive way to improve the situation could be to simply modify the *-product in order to cut off the nonlocal phenomena at large enough distances.
For this aim it is useful to write the *-product in the integral representation,
\begin{eqnarray}
f(x)*_W g(x)=\int d^D z\;d^D y \; \frac{1}{\pi^D \det \theta} \exp[-2i(x\theta^{-1}y+y\theta^{-1}z+z\theta^{-1}x)] f(y) g(z),\nonumber \\
\label{}
\end{eqnarray}
where $W$ indicates the Moyal *-product. The integral gets contributions from $f(y)$ and $g(z)$ for all values of $y$ and $z$.
Now different ways to impose a cutoff to the integral can be considered, resulting in different modified *-products. 
A Gaussian cutoff gives (restricting to a plane for simplicity)
\begin{eqnarray}
f(x)*'g(x)&:=\int d^2 z\;d^2 y\;\frac{1}{\pi^2 \det \theta} \exp[\frac{i}{\theta}(x\wedge y+y\wedge z+z\wedge x)] \nonumber\\
&\exp[\frac{-1}{\theta}\left((x-y)^2+(x-z)^2\right)] \; f(y) \; g(z),
\label{integral}
\end{eqnarray}
and the step function leads to
\begin{eqnarray}
f(x)*''g(x)&:=\int d^2 z\;d^2 y\;\frac{1}{\pi^2 \det \theta} \exp[\frac{i}{\theta}(x\wedge y+y\wedge z+z\wedge x)] \nonumber\\
&\Theta(l^2-(x-y)^2) \; \Theta(l^2-(x-z)^2) \; f(y) \; g(z).
\label{step1}
\end{eqnarray}
For the $*'$-product one can check explicitly that it satisfies the properties
\begin{eqnarray}
[x_1,x_2]_{*'}=i\theta,
\label{}
\end{eqnarray}
and
\begin{eqnarray}
e^{ixk}*'e^{ixq}=e^{ix(k+q)} e^{\frac{-i\theta}{2}k\wedge q} e^{\frac{-\theta}{4}(k^2+q^2)}.
\label{exp}
\end{eqnarray}
Thus it realizes the proper commutator for NC coordinates and provides a new factor in the product of two plane waves that could
act as a cutoff for UV-physics. However, from (\ref{exp}) one can easily deduce that the product is not associative and thus it can not be used to replace the Moyal *-product in NC field theory.
If nonassociativity can be dealt with in the field theory, the *-products with cutoff could be used to study the physical
implications of finite nonlocality.

An alternative approach to finite range NC was developed in \cite{Bahns:2006cp}. The authors considered
a noncommutative space-time where the commutator of coordinates of two distinct points have a compact support, vanishing if the points are far apart.
This leads to a *-product which reduces to the ordinary product for fields in points whose separation is outside the support of $\theta$.
The authors were unable to construct an interacting field theory in this approach, and further development in this direction is still lacking.

Finally, we note that the nonlocal effects in NC field theory can be completely removed by allowing the parameter of noncommutativity to be composed of fermionic parameters \cite{Gitman:2007jp}. If the parameter of noncommutativity is taken in the bifermionic form, $\theta^{\mu\nu}=i\theta^\mu\theta^\nu$, where $\theta^\mu$ are Grassmann odd, the series expansion of the *-product terminates at finite order. 
In this framework the modifications to the commutative model are rather mild and renormalizability properties are improved compared to the usual NC theories \cite{Fresneda:2008sr}.  This formulation provides an interesting version of noncommutativity to be studied further.


%% file: chap02.tex
\chapter{Symmetries of NC space-time and general properties of NC QFT}


\section{Symmetries of constant $\theta^{\mu\nu}$}
The commutation relation (\ref{commutator}) provides a structure that has to be preserved under symmetry transformations of the space-time.
Thus the symmetry of the NC space-time is given by the subgroup of the Poincar\'e group that is the stability group of the antisymmetric tensor $\theta^{\mu\nu}$.
  The stability group that preserves
 the commutation relation in four dimensional space-time
is given by $SO(1,1)\times SO(2)$ combined with translations \cite{AlvarezGaume:2001ka, AlvarezGaume:2003mb} and the physical effect of $\theta^{\mu\nu}$ is similar to a background field that provides preferred directions in space-time.

The symmetry of space-time has profound consequences in field theory, since the particles are
representations of the space-time symmetry group. Violation of Lorentz symmetry also leads to important
 phenomenological effects that provide ways to detect the physical consequences of noncommutativity experimentally. 
Such Lorentz-violating effects could be observed for example as diurnal variation of scattering 
cross sections and polarization dependent speed of light \cite{ Hewett:2000zp,Godfrey:2001yy,Baek:2001ty,Grosse:2001xz,Guralnik:2001ax,
Chaichian:2000si,Carroll:2001ws,Mocioiu:2000ip,Anisimov:2001zc,Carlson:2001sw}.
From the lack of evidence for such phenomena, bounds on the scale of noncommutativity can be derived. 
Especially, the vacuum birefringence phenomenon poses very restrictive bounds on $\theta$ when compared with
cosmological observations unless supersymmetry arguments are invoked; see \cite{Jaeckel:2005wt} and {\bf I}. 

A different point of view to space-time symmetries in NC quantum field theory is obtained by considering the symmetries in the context of quantum groups.
It was realized in \cite{Chaichian:2004za} that noncommutative space-time with the Moyal *-product 
is symmetric under the action of the twisted Poincar\'e symmetry that is obtained by twisting the Hopf algebra structure 
of the Lie algebra of the ordinary Poincar\'e group.  

In the standard case the Lie algebra structure of Poincar\'e transformations is given by the commutators
\begin{eqnarray}
[P_\mu ,P_\nu ]                  			&=& 0,  \nonumber \\
\left[ M_{\mu\nu} , M_{\alpha\beta} \right] &=& -i ( \eta_{\mu\alpha} M_{\nu\beta} - \eta_{\mu\beta} M_{\nu\alpha} - \eta_{\nu\alpha} M_{\mu\beta} + \eta_{\nu\beta} M_{\mu\alpha} ) , \label{}\\
\left[ M_{\mu\nu} , P_\alpha        \right] &=& -i ( \eta_{\mu\alpha} P_\nu - \eta_{\nu\alpha} P_\mu ) , \nonumber
\end{eqnarray}
and the coproduct for all generators is given by
\begin{eqnarray}
\Delta_0(Y) = Y \otimes  1 + 1 \otimes  Y.
\label{}
\end{eqnarray}
The idea behind the twist is to change the coproduct while leaving the algebra itself unchanged.
Choosing the Abelian twist 
\begin{eqnarray}
\mathcal{F}= \exp{\left( \frac{i}{2}\theta^{\mu\nu} P_\mu \otimes P_\nu \right)},
\label{}
\end{eqnarray}
one obtains for the generators of the Lorentz algebra the deformed coproduct
\begin{eqnarray}
\Delta_\theta(M_{\mu\nu}) &:=&  \mathcal{F} \Delta_0(M_{\mu\nu}) \mathcal{F}^{-1}\nonumber \\
					&=&   M_{\mu\nu} \otimes  1 + 1 \otimes  M_{\mu\nu} -\frac{1}{2}\theta^{\alpha\beta} [(\eta_{\alpha\mu} P_\nu - \eta_{\alpha\nu} P_{\mu}) \otimes  P_\beta  \label{}\\
					&+&  P_\alpha \otimes  (\eta _{\beta\mu} P\nu - \eta_{\beta\nu} P_\mu )]. \nonumber
\end{eqnarray}
Due to the commutativity of the $P_\mu$'s the coproduct of the translation generators is left unchanged by the twist.
Basically, the Hopf algebra structure of a group defines how the group acts on products of representations. If the coproduct is changed,
it has implications also on the product of representations. Consider as the representation space the algebra of functions in Minkowski space, with
\begin{eqnarray}
P_\mu f(x)     &=& i\partial_\mu f(x), \nonumber\\
M_{\mu\nu}f(x) &=& i(x_\mu \partial_\nu -x_\nu \partial_\mu)f(x). \nonumber
\end{eqnarray}
In the case of ordinary Poincar\'e Hopf-algebra the product in the representation algebra is the commutative pointwise product:
\begin{eqnarray}
m_0 (f\otimes g)(x) =f(x)g(x).
\label{}
\end{eqnarray}
Then the new product for the representation of the twisted Hopf algebra is obtained from the requirement
\begin{eqnarray}
Y \triangleright m_\theta (f\otimes g) (x)= m_\theta (\Delta_\theta (Y) (f\otimes g)) (x),
\label{leibniz}
\end{eqnarray}
where $Y\in \mathcal{P}$, the Poincar\'e algebra.
The new associative product $m_\theta$ is defined by
\begin{eqnarray}
m_\theta(f(x)\otimes g(x))=m_0 \circ \mathcal{F}^{-1} (f \otimes g) (x),
\label{}
\end{eqnarray}
which guarantees that it satisfies the Leibniz rule (\ref{leibniz}).
The twisted product $m_\theta$ is exactly the Moyal *-product (\ref{star}).
An important feature of the twisted Poincar\'e transformations is that they leave the commutator 
$[x^\mu , x^\nu]$ invariant, thus keeping $\theta^{\mu\nu}$ nontransforming, and in this sense the twisted Poincar\'e algebra
is a symmetry of the NC space-time.

An important consequence follows from the fact that the representation content of the twisted symmetry group is exactly identical with 
the non-twisted group. Thus, even though
ordinary Poincar\'e symmetry is not a symmetry of the theory, all particles should sit on the representations of the Poincar\'e group,
 justifying the use of the usual representations that has been widely adopted in the study of noncommutative field theories.
Thus the twisted Poincar\'e algebra offers a firm
framework for the proofs of the NC version of results such as the CPT, spin-statistics
and the Haag theorems \cite{Chaichian:2004yh}.

It should be noted that covariance of the *-product can be obtained alternatively by considering coordinate transformations under which also $\theta$ transforms.  For references and comparison with the twist-approach, see \cite{GraciaBondia:2006yj}.

On what comes to the possible further implications of the twisted space-time symmetry, 
there seems to be controversy in the field. In \cite{Balachandran:2005eb,Balachandran:2005pn} the authors argue that also the product of the fields' Fourier modes should be affected by the twist,
resulting in a modified oscillator algebra. One implication of this approach is that the action of a NC field theory
reduces to the commutative action. Then the effect of noncommutativity is seen only in the modified statistics of fields \cite{Balachandran:2005pn}.
These results were however argued to be false in \cite{Tureanu:2006pb, Chaichian:2008ge}. 
We conclude the review of the twist deformed symmetry by noting that the
twisted space-time symmetry in NC field theory is an active area of study where final conclusive understanding seems to be still lacking.
For a recent survey on these issues see \cite{Chaichian:2008ge}


\subsection{Curved NC space-time and NC gravity}


As the very idea of noncommutativity stems from the deeper structure of space-time, presumably relevant at Planck's scale, it is natural
to try to implement also gravitational physics and general relativity into the NC QFT setting. A large amount of work
can be found in the literature approaching the problem from different points of view \cite{Chamseddine:2000zu, Rivelles:2002ez, Harikumar:2006xf, Steinacker:2007dq, Grosse:2008xr, Aschieri:2005yw, 
Calmet:2005qm, Kobakhidze:2006kb, Chaichian:2008pq, Marculescu:2008gw}.
It is notable that the noncommutative gauge transformations are related to space-time transformations implying a possible deep connection
between NC gauge symmetry and gravity. In \cite{Rivelles:2002ez}
it was noted that the noncommutative analogue of $U(1)$ gauge theory can be interpreted as a commutative U(1) gauge theory coupled to gravity, where the curved metric
arises from noncommutativity. This type of emergent gravity theories have been further studied in  \cite{Steinacker:2007dq, Grosse:2008xr} 
relating the gravity theory to matrix model formulation of noncommutative gauge theory.

Other attempts to generalize the NC field theories from flat NC space-time to a more general case have been considered in the literature.
An obvious way towards general covariant field theory action is provided by replacing the space-time derivatives in the
*-products by covariant derivatives \cite{Harikumar:2006xf}. This replacement however leads to nonassociative *-product.

In \cite{Calmet:2005qm} the authors noted that $\theta^{\mu\nu}$ is left invariant under a special subset of general coordinate transformations.
This invariance allows one to construct the unimodular version of gravity in NC space-time. In \cite{Calmet:2005qm} the action for NC unimodular gravity
was computed up to second order in $\theta$, using the Seiberg-Witten map.
More recently, a noncommutative gravity based on gauging the noncommutative analogue of the $SO(1,3)$ symmetry was developed in \cite{Marculescu:2008gw}. 

Approaches towards NC gravity theory using the idea of twisted symmetry have been developed in \cite{Aschieri:2005yw, Chaichian:2008pq}.
Generalization of the twisted Poincar\'e symmetry to twisted diffeomorphisms lead the authors of \cite{Aschieri:2005yw} to a NC general relativity theory.
This approach, however, has been criticized as it is not consistent with the most strict formulation of the gauge principle \cite{Chaichian:2006we}.
In   \cite{Chaichian:2008pq}  a similar approach was taken, but now with a consistent implementation of the gauge principle.
This type of covariantization of the twist leads to nonassociativity as can be anticipated from the results in \cite{Harikumar:2006xf}.


\section{Lorentz-invariant formulation of NC space-time}

An alternative to the canonical formulation with the constant parameter of noncommutativity is to consider $\theta^{\mu\nu}$ as
an element of the algebra that transforms as a Lorentz tensor. Then the possible values of $\theta^{\mu\nu}$ provide six degrees
of freedom in addition to the four coordinates $x^\mu$. This approach is similar to the one taken in the seminal paper of Doplicher, Fredenhagen and Roberts \cite{Doplicher:1994zv}.
Now the structure of the NC space-time is given by the DFR-algebra
\begin{eqnarray}
\left[\hat x^\mu , \hat{x}^\nu\right]=i\hat\theta^{\mu\nu},\nonumber\\ 
\left[\hat x^\mu , \hat \theta^{\nu\rho} \right]=0, \label{algebra}\\ 
\left[\hat \theta^{\mu\nu} , \hat\theta^{\rho\sigma} \right]=0\nonumber,
\end{eqnarray}
The last equation follows from the Jacobi identity and the other two equations.

Due to the commutativity of $\theta^{\mu\nu}$, the algebra of noncommutative fields can be realized using the Moyal *-product (\ref{star}) just as in the case
of canonical noncommutativity.
Now choosing a state where the operator $\theta^{\mu\nu}$ takes an exact value corresponds to the usual
Moyal space-time  or, in the case of $\theta^{\mu\nu}=0$, to the commutative space-time. A more general state
leads to a range of different values for $\theta^{\mu\nu}$
and the trace of noncommutative field operators is now given by a space-time integral of their *-products appended
with an integral over the values of $\theta$ corresponding to the chosen state.
Although the approach is completely Lorentz covariant, Lorentz invariance is lost by choosing a state that 
optimizes the uncertainty relations arising from (\ref{algebra}).
Since the space of the possible values of $\theta^{\mu\nu}$ that satisfy the uncertainty relations proposed in \cite{Doplicher:1994zv} does not
allow for a Lorentz invariant average, the best one could do is a rotation invariant theory.

The step proposed by Carlson, Carone and Zobin \cite{Carlson:2002wj} was to extend the integral to all values of $\theta^{\mu\nu}$ in $\mathbb{R}^6$ and put
the details of the state into an unknown weight function $W(\theta)$ \footnote{Originally Carlson et al. derived their theory from the NC space-time of Snyder \cite{snyder}.
A restatement of their construction in terms of the DFR-algebra was given in \cite{Morita:2002cv}.}
\footnote{An explicit example of the weight function is introduced in \cite{kwee}.}:

\begin{eqnarray}
Tr \hat \phi (\hat x , \hat \theta) = \int d^4 x\;d^6\theta\; W(\theta) ~ \phi(x,\theta).
\label{}
\end{eqnarray}
If $W$ is chosen to be a Lorentz scalar, this construction leads to Lorentz-invariant noncommutative field theory.
One important consequence is the absence of Lorentz violating phenomena such as vacuum birefringence. This allows for the scale of noncommutativity
to be relatively low and noncommutative effects could be significant already much below the Planck scale.
For example, an experimental bound on  $\theta$ arising from Lorentz-invariant NC corrections to Bhabha scattering, dilepton and diphoton production leads to the rather permissive bound \cite{Conroy:2003qn}
\begin{eqnarray}
\Lambda_{NC} \sim \sqrt{\langle\theta^2\rangle}  >  160\;Gev\;\;\;\; 95\%\; \text{C.L..}
\label{}
\end{eqnarray}
As was shown in {\bf III} the Lorentz-invariant theories have significant problems related to unitarity and causality. These issues are explained in the rest of this chapter.


\section{Unitarity }

From the start, the foundational basic properties of NC quantum field theories such as unitarity and  causality have been under extensive study in the literature, 
both in the axiomatic field theory approach and in more practical calculations in specific models.
Due to the lack of Lorentz invariance, theories with different values of $\theta^{\mu\nu}$ can have very different properties.
Especially, time-space noncommutativity seems to induce various difficulties in NC theories, that are absent in the case of vanishing $\theta^{0i}$. 

\subsection{Unitarity  in the canonical case}
\begin{figure}
\begin{center}
\includegraphics[height=3cm]{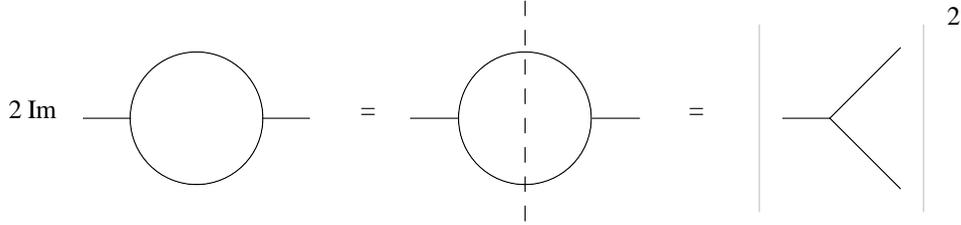}
\caption{The cutting rule}
\label{Optical}
\end{center}
\end{figure}

It has been realized that unitarity is violated in quantum field theory if time and space do not commute \cite{Gomis:2000zz} while such problem does not
generally appear if only $\theta^{ij}$ components are allowed to be nonzero. 
In \cite{Gomis:2000zz} the optical theorem was checked.
For a one loop two-point function in $\phi_*^3$ theory the optical theorem is expressed by the cutting rule given in Fig. \ref{Optical}.
It was shown that the cutting rule is satisfied provided
\begin{eqnarray}
p\circ p := -p^\mu \theta_{\mu\nu}^2p^\nu > 0.
\label{}
\end{eqnarray}
In Minkowski space the inequality is satisfied in general only if $\theta_{0i}=0$, implying violation of
unitarity in theories with noncommuting time.

Unlike theories with $\theta_{0i}=0$, field theories with noncommutative time can not be obtained as an approximate description
of a limit of string theory. The $\theta^{0i}\not=0$ case is obtained in string theory in the presence of a background electric field.
However, while it is possible to find a limit with nonvanishing $\theta^{0i}$ where the closed strings decouple, 
it is not possible  to decouple  massive open string states while keeping $\theta^{0i}$ finite \cite{Seiberg:2000gc, Gopakumar:2000na}.
Therefore, unitarity of the corresponding string theory can not be used as an argument for the unitarity of the $\theta^{0i}\not=0$ QFT. 
 
On the other hand, starting from a Hermitian Lagrangian one should obtain a unitary perturbation theory through the standard quantization procedures.
 In fact it was realized soon that in the case of noncommuting time, the time-ordered perturbation theory (TOPT) that one obtains in the Hamiltonian 
formulation of perturbation theory does not reduce to the covariant perturbation theory with the noncommutative analogues of the usual Feynman rules \cite{Liao:2002xc, Bahns:2002vm}.
The time-ordering in the $S$-matrix
\begin{eqnarray}
S=T \exp\left({i\int d^4x \; \mathcal{L}_{int} }\right)
\label{}
\end{eqnarray} 
leads initially to graphs with specific time-ordering of the vertices.
In the usual QFT the time-ordered diagrams can be rewritten in a Lorentz-covariant form by combining the time-ordered propagators to covariant Feynman propagators.
Now if there are time derivatives in the *-products inside the interaction Hamiltonian, the time-ordering clashes with the *-product in such a way that the usual covariantization is not possible.

\begin{figure}
\begin{center}
\includegraphics[height=3cm]{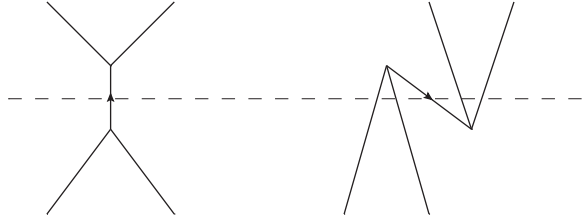}
\caption{TOPT diagrams}\label{TOPT}
\end{center}
\end{figure}

This explains why the naive use of covariant perturbation theory can lead to nonunitary results in the case of noncommuting time, as it is not equivalent to the manifestly unitary 
Hamiltonian formulation of perturbation theory. However, even if one works with the TOPT, 
one encounters problems with unitarity when considering gauge theories \cite{Ohl:2003zd}.
Thus resorting to the time-ordered perturbation theory can not be considered as a solution to the unitarity problem.

\subsection{Unitarity in the Lorentz-invariant case}
Perturbative unitarity in the case of the Lorentz-invariant formulation of NC field theory has been studied in  \cite{Morita:2003vt} and {\bf III}.
In \cite{Morita:2003vt} the unitarity issue was addressed by studying the optical theorem in a one-loop calculation in $\phi_*^4$-theory. 
The calculation was done in the covariant perturbation theory approach and the result confirmed the unitarity in such theory at one loop level.

This result leads to the question whether time-ordered and covariant perturbation theories could be equivalent in the Lorentz-invariant NC QFT.
This, however, turns out not to be true. In {\bf III} a simple tree-level amplitude in a Lorentz-invariant NC scalar field theory, 
described by the action
\begin{eqnarray}
S=\int d^6\theta d^4 x\;W(\theta) \left(  \frac{1}{2}(\partial_\mu \phi)^2-\frac{m^2}{2}\phi^2  - \frac{\lambda}{3!}\phi*\phi*\phi \right),
\label{}
\end{eqnarray}
was calculated both in TOPT and in covariant perturbation theory.
The TOPT amplitude obtained from the sum of the time-ordered diagrams, depicted in Fig. 3.2, turns out to be 
inequivalent to the amplitude obtained from covariant perturbation theory. The discrepancy is due to the contributions arising upon integration over the $\theta^{0i}$ components; if $\theta^{0i}$ were kept fixed to 0, the TOPT and covariant expressions would coincide.
Thus the Lorentz invariant integration over $\theta$ does not cancel the harmful contributions arising from time-space components of $\theta$.

Due to this result, the generalization of unitarity to a general Lorentz-invariant NC field theory seems doubtful
and thus one should rely on the time-ordered formulation in order to obtain manifestly unitary perturbation theory.  
However, as already mentioned, even in the TOPT approach unitarity seems to be lost in physical theories that include gauge fields.


\section{Causality}

In a Lorentz invariant theory the requirement of causality means that all events that are causally connected preserve their time-ordering
in all inertial frames. In other words, only time-like or light-like separated events are allowed to be causally connected.
In quantum field theory, this is often expressed in terms of {\it microscopical causality condition} which means
that the commutator of local densities of observables can be nonvanishing only if they are evaluated in points that are time-like or light-like separated, i.e. one point
resides inside or on the light-cone originating from the other point.
The inherent nonlocality in noncommutative theories presumably brings modifications
to the causality properties of field theory and will be discussed in this section.


\subsection{Causality in the canonical case}

As the Lorentz symmetry is reduced to $SO(1,1)\times SO(2)$ in the presence of nonzero constant $\theta^{ij}$, the light-cone
causality condition is modified to the light wedge which is invariant under the symmetry group.    
This can be seen by considering the commutator of two observables \cite{Chaichian:2002vw,Chu:2005nb}
\begin{eqnarray}
C=[O(x),O(y)].
\label{}
\end{eqnarray}
Observables (more strictly speaking, the local densities of observables) are in 
general constructed by taking local products of fields. In the noncommutative theory this corresponds
 to *-products of fields. As a simple example observable one can consider e.g. 
$O(x)=:\phi(x)*\phi(x):$, where the normal ordering is assumed in order to simplify the calculation.
The *-product in the definition of the local observable is a source for nonlocality even if $\phi(x)$
is taken to be a free field.
 
For our purposes it is enough to consider a single matrix element of the operator, e.g.
\begin{eqnarray}
M=\bra{0}\left[O(x),O(y)\right]\ket{p,p'}.
\label{comm}
\end{eqnarray}  
To evaluate (\ref{comm}) one simply inserts the standard expansion of the free scalar field in terms of creation and annihilation operators.
This results in \cite{Chaichian:2002vw}
\begin{eqnarray}
\mathcal{M}&=& -\frac{2i}{(2\pi)^6}\frac{1}{\sqrt{(\omega_\mathbf{p}\omega_{\mathbf{p}'})}}(e^{-ip'x-ipy}+e^{-ipx-ip'y})\nonumber\\
&\times&\int\frac{d^3k}{\omega_\mathbf{k}} \sin[\mathbf{k}(\mathbf{x}-\mathbf{y})]\;\cos\left(\frac{1}{2} k\wedge p \right)\;\cos\left(\frac{1}{2} k\wedge p'\right).
\label{}
\end{eqnarray}
Here $\omega_\mathbf{k}=\sqrt{\mathbf{k}^2+m^2}$. 
Obviously the r.h.s. is nonzero only when $\theta^{0i}\not=0$, implying loss of causality in the presence of time-space noncommutativity. By the symmetry of the $\theta^{0i}=0$ case,
the causality condition is given by the light wedge when time and space commute. 
Similar result was obtained also in a different approach in \cite{Chu:2005nb}, where the commutator
of Heisenberg fields in two points was considered. Then noncommutative effects arise only when interactions are taken into account,
and the light wedge was shown to arise in a perturbative one-loop calculation. 

\begin{figure}
\begin{center}
\includegraphics[height=5cm]{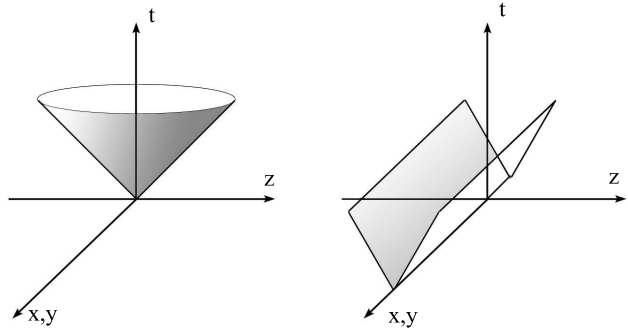}
\caption{Light cone and light wedge}
\label{causalityfigure}
\end{center}
\end{figure}


\subsection{Causality in the Lorentz-invariant case}

In the Lorentz invariant case 
the causality condition, ie. the domain of validity of the equation 
\begin{eqnarray}
\left[O(x),O(y)\right]=0,
\label{}
\end{eqnarray}
should be Lorentz invariant. Thus one may entertain the hope of a light-cone causality condition. However, the integration over $\theta$ 
in the noncommutative action brings contributions from all values of the noncommutativity parameter and thus one might expect
nonlocality to spread infinitely in all directions. The causality issue was investigated in {\bf III} by calculating the commutator 
of two observables located in different space-time points following the analysis of \cite{Chaichian:2002vw}.
In the Lorentz invariant theory the local observables should have no Lorentz-violating $\theta$-dependence and thus they 
should also include the $\theta$-integration:
\begin{eqnarray}
O(x)=\int d^6  \theta  \; W(\theta) :\phi(x)*\phi(x):
\label{}
\end{eqnarray}
The result,
\begin{eqnarray}
\mathcal{M}_\mathrm{LI}&=&-\frac{2i}{(2\pi)^6}\frac{1}{\sqrt{(\omega_\mathbf{p}\omega_\mathbf{p}')}}(e^{-ip'x-ipy}+e^{-ipx-ip'y}) \nonumber\\
&\times&\int\frac{d^3k}{\omega_\mathbf{k}} \left\{\sin[\mathbf{k}(\mathbf{x}-\mathbf{y})]\; \int d\theta_1 W(\theta_1)\cos\left(\frac{1}{2} k\wedge_1 p \right) \right. \label{causalityint} \\
&\times&\left. \int d\theta_2 W(\theta_2) \cos\left(\frac{1}{2}k \wedge_2 p'\right)\right\},
\nonumber
\end{eqnarray} 
exhibits the infinite nonlocality of the theory as the matrix element is in general nonvanishing for all $(x,y)$ for
 reasonable weight functions $W(\theta)$, including the Gaussian. 
 
The mixing between short and long distance degrees
of freedom can be understood in terms of the modified causality condition. Due to infinite
propagation speed all distances can be correlated and as we
have seen, in the Lorentz-invariant case causality is violated in all
directions and thus the UV/IR mixing is expected to appear. Choosing $W$ in the Gaussian form, the $\theta$-integration effectively replaces the oscillating phase factors in loop diagrams by Lorentz-invariant Gaussian damping factors \cite{Morita:2002cv, Morita:2003vt}:
\begin{eqnarray}
\int d\theta \; W(\theta) \; e^{-ik\wedge p}\longrightarrow e^{-a^4(k^2 p^2 - (k\cdot p)^2)/4},
\label{}
\end{eqnarray}
where $a$ is a parameter describing the scale of noncommutativity.
Despite this modification, IR singularity still arises as the external momentum goes
to zero. However, it was argued in \cite{Morita:2002cv} that the problem may be
avoided by a suitably chosen IR limit under which $a$ goes to zero
with the external momentum. In any case, lack of causality is a problem
that cannot be dismissed and it is necessary to find a way to
restore the light-cone causality in order to achieve a consistent Lorentz-invariant NC
field theory.


\section{Exact NC QFT}

Most of the study of NC QFT has been done in the Lagrangean approach. Attempts to generalize the results of axiomatic approach
to quantum field theory have been also made in the literature. In the canonical NC case with constant $\theta^{\mu\nu}$, some of the postulates of the usual axiomatic QFT, such as Lorentz
symmetry have to be replaced by other postulates.
The first step towards axiomatic NC QFT was taken in \cite{AlvarezGaume:2003mb}, where the usual Wightman functions were investigated taking
$O(1,1)\times SO(2)$ as the symmetry group, and the validity of the CPT theorem was shown. 
From the point of view of the reconstruction theorem \cite{wightman} this formulation would lead to a 
ordinary QFT with the symmetry group $\mathcal{P}(1,1)\times E_2$.
In order to make the axiomatic formulation genuinely noncommutative, in \cite{Chaichian:2004qk} an alternative definition
for the NC Wightman functions was proposed with generalized *-products inserted between the fields:
\begin{eqnarray}
W_*(x_1,x_2,...,x_n) &=& \bra{0}\phi(x_1)*\phi(x_2)*...*\phi(x_n)\ket{0}, \nonumber \\
\phi(x)*\phi(y)      &:=&  \phi(x)   e^{    \frac{i}{2} \theta^{\mu\nu} \frac{\overleftarrow\partial }{\partial x^\mu} \frac{\overrightarrow\partial }{\partial y^\nu}   } \phi(y).
\label{}
\end{eqnarray}

With the $W_*$ several noncommutative analogues of familiar results of axiomatic QFT have been proven.
The CPT and spin-statistics theorems were proven in \cite{Chaichian:2004qk}. Analytical properties of scattering amplitudes
were studied in \cite{Chaichian:2004hb,Tureanu:2006ct} and it was shown in \cite{Chaichian:2004hb}
the results are sensitive to the form of the causality condition. If the causality condition is taken to be defined by the light wedge, 
the NC theory suffers from a severe lack of analyticity and reduces the predictive power of the theory.
On the other hand,  a causality condition of the form
\begin{eqnarray}
x_0^2-x_3^2-x_2^2-x_1^2<-l^2,
\label{}
\end{eqnarray}
that exhibits nonlocality of finite range, is actually equivalent to the usual causality condition of the commutative theory
according to an old result of Wightman-Vladimirov-Petrina \cite{WVP}. In this case analyticity properties similar to the commutative case
would be obtained as argued in \cite{Chaichian:2004hb}.
We emphasize that the causality condition obtained in the Lorentz-invariant NC QFT studied in {\bf III} does not fall into this category as there the nonlocality
is infinite. To obtain a causality condition with finite nonlocality a new approach is needed as discussed in Section 2.3.

%% file: chap03.tex
\chapter{NC gauge theories and model building}

Due to the local nature of gauge transformations, noncommutativity
introduces profound changes in NC gauge field theories. These changes and their consequences in model building
have been an intensive area of study during the recent years. In this chapter various approaches to treating
NC gauge symmetries and subsequent construction of particle physics models based on NC QFT are reviewed.


\section{Gauge theories in NC space-time}


\subsection{The NC no-go theorem.}

The gauge transformations in noncommutative space-time are affected by the nonlocal product which leads to new interesting properties.
In the *-product formalism the product of two matrix-valued gauge transfromations is given by the matrix product combined with the *-product,
\begin{eqnarray}
(U(x),V(x))\mapsto U_i^j(x)*V_j^k(x),
\label{stargauge}
\end{eqnarray}
and they act on the fields via
\begin{eqnarray}
(U,\phi)\mapsto U_i^j(x)*\phi_j(x).
\label{}
\end{eqnarray}
From the transformation law one can deduce several restrictions on the possible choices for the gauge groups and their representations. An obvious consequence of the *-product is that the NC analogue of $U(1)$ gauge transformations becomes 
noncommutative, resembling in this sense the nonabelian theories in commutative space-time.
In \cite{Hayakawa:1999yt} the noncommutative analogue  $U(1)$ gauge field theory was considered and it was observed that
noncommutative QED has the intriguing property that the electric charge is quantized to the values $\pm 1$ and 0.
The restrictions arising from the transformation law (\ref{stargauge}) were thoroughly analyzed in \cite{Chaichian:2001mu} and the results were gathered under a no-go theorem:
\begin{enumerate}
\item A straightforward noncommutative generalization is possible only for the unitary group. For example, the *-product of two $SU(N)$ matrices does not close in general to $SU(N)$. 
\item The only allowed representations are fundamental, antifundamental, bifundamental, adjoint, and the trivial representations, i.e. no higher rank tensor representations are possible. A field can be charged at most under two gauge groups, being in the fundamental representation of one group and in the antifundamental representation of the other group.
\end{enumerate}

These restrictions, if not circumvented somehow, lead to severe consequences on model building in noncommutative space-time.
For example, the fractional charges of quarks seem to be inconsistent with this theorem. Also higher rank representations are required
for the matter content of Grand Unified Theories, but prohibited by the theorem.
Thus a way to circumvent the implications of the theorem seems to be necessary for noncommutative model building.

Formulation of NC gauge theory in a way that allows the use of ordinary gauge transformations while still
maintaining the noncommutative products between fields in the Lagrangian can be obtained by defining {\it twisted gauge symmetries} similarly
to the twisted space-time transformations \cite{Vassilevich:2006tc, Aschieri:2006ye}. 
As was argued in \cite{Chaichian:2006we}, this approach seems to be inconsistent with the most strict formulation of the gauge principle and a more consistent 
formulation is obtained by covariantizing the derivatives in the *-product \cite{Chaichian:2006wt}.
However, as in the case of gravity theory with covariant twist, the gauge-covariant twist leads to a nonassociative *-product.
In the rest of this section we review two alternative approaches for dealing with the restrictions of the no-go theorem. 


\subsection{The Seiberg-Witten map}
The noncommutative low-energy gauge field theory obtained from string theory in a background field can be
written in terms of a commutative gauge field theory with a $\theta$-dependent action by choosing an alternative regularization \cite{Seiberg:1999vs}. The correspondence
between these alternative descriptions is given by the Seiberg-Witten map, which provides a way to write
a noncommutative gauge field theory in terms of an ordinary gauge field theory with $\theta$-dependent action. 
This idea was further developed in \cite{Jurco:2000ja,Madore:2000qh,Jurco:2001rq} where it was shown that any commutative (but possibly nonabelian) gauge algebra can be extended to a noncommutative version by a generalization of the Seiberg-Witten map. This way the Seiberg-Witten map provides a way to write
a noncommutative gauge field theory in terms of an ordinary gauge field theory with a $\theta$-dependent action. 

To obtain the Seiberg-Witten map between a commutative $su(n)$ gauge field theory and its noncommutative analogue, consider NC fields $\hat A_\mu$ and $\hat \phi$ whose
gauge transformations are generated by the gauge parameter $\hat \Lambda$ which takes values in the enveloping algebra of $su(n)$:
\begin{eqnarray}
\hat \delta \hat A_\mu &=& \partial_\mu\hat \Lambda + i[\hat \Lambda, \hat A_\mu]_*,	\\
\hat \delta \hat \phi  &=& i\hat \Lambda * \hat \phi. 			
\label{}
\end{eqnarray}
The noncommutative fields and gauge parameter are to be written as functions of the commutative variables,
\begin{eqnarray}
\hat \Lambda &=& \Lambda + \hat\Lambda'[\Lambda, A_\nu, \theta], \\
\hat A_\mu   &=& A_\mu   + \hat A'_\mu [ A_\nu, \theta], \\ 
\hat \phi    &=& \phi    + \hat \phi'  [\phi, A_\nu, \theta],
\label{}
\end{eqnarray}
which transform as ordinary $su(n)$ fields:
\begin{eqnarray}
\delta A    &=& \partial_\mu \Lambda + i [\Lambda, A_\mu], \\
\delta \phi &=& i\Lambda\phi.
\label{}
\end{eqnarray}
Then the unknown expressions $A'$, $\Lambda'$ and $\phi'$ are determined by requiring 
equivalence between the noncommutative and commutative gauge transformations \cite{Seiberg:1999vs},
\begin{eqnarray}
\hat A_\mu + \hat \delta \hat A_\mu &=& \hat A_\mu[A_\nu+\delta A_\nu, \theta], \\
\hat \phi  + \hat \delta \hat \phi  &=& \hat \phi [\phi, A_\nu+\delta A_\nu, \theta],
\label{}
\end{eqnarray}
and can be solved perturbatively. In the lowest order:
\begin{eqnarray}
\hat A_\mu   &=& A_\mu   + \frac{1}{4} \theta^{\rho\sigma} \left\{A_\sigma , \partial_\rho A_\mu+ F_{\rho\mu} \right\}, \\
\hat \phi    &=& \phi    + i\theta^{\rho\sigma} A_\sigma \partial_\rho \phi +\frac{i}{8}\theta{\rho\sigma} [A_\rho , A_\sigma]\phi, \\
\hat \Lambda &=& \Lambda + \frac{1}{4}\theta^{\rho\sigma}\left\{ A_\sigma , \partial_\rho \Lambda \right\}.
\label{}
\end{eqnarray}

Inserting the expansion into the Lagrangian and expanding the *-products in $\theta$, one obtains the noncommutative theory Lagrangian as one with commutative gauge fields
and $\theta$-dependent correction terms. The Seiberg-Witten map provides a way to write noncommutative gauge field theory in terms of an ordinary gauge field theory
and can be used to build noncommutative models by mapping them to ordinary commutative gauge field theories with noncommutative correction terms.

 
\subsection{Modified gauge transformations}

To circumvent the restrictions in noncommutative model building without resorting to the Seiberg-Witten map which requires one to define the noncommutative
theory in terms of commutative concepts, a modified version of gauge transformation in NC space-time was introduced in \cite{Chu:2001kq}.
An important ingredient in this construction is the half-infinite NC Wilson line operator for the $U_*(n)$- gauge group:
\begin{eqnarray}
 W_C(x)&=&P_*\exp\left(ig\int_0^1
 d\sigma {d\z^\mu(\sigma) \over d\sigma}A_\mu(x+\z(\sigma))\right)\,, \label{wilson}
\end{eqnarray}
where the integration is along the contour $C$ from $\infty$ to $x$,
\begin{eqnarray}
 C=\left\{\z(\sigma)\,,0\le \sigma \le 1\,|\,\z(0)=\infty\,,\z(1)=0\right\}\,,
 \label{path}
\end{eqnarray}
and the path ordering involves the Moyal *-product between all functions.
Assuming that the gauge transformation goes to unity at infinity, the transformation law for the half-infinite
Wilson line reads:
\begin{eqnarray}
 W_C(x)\rightarrow W_C(x)*U^{-1}(x)\,.
\end{eqnarray}
Now we can define a two-index tensor field by the transformation law:
\begin{eqnarray}
\phi(x)\longrightarrow \phi^U=(1\otimes U*W^{-1})*(U\otimes W)*\phi.
\label{modified}
\end{eqnarray}
This transformation law is closed and thus provides a consistent definition for a two-index representation of the gauge transformation. 
Without the *-products, i.e. in the commutative limit, this transformation reduces to ordinary commutative transformation of two-index field.
A notable difference to the commutative tensor representation is that this transformation law does not commute with the exchange of the
tensor indices, $\phi^{ij}\rightarrow \phi^{ji}$, and thus it does not reduce to symmetric and antisymmetric representations.
A generalization of this type of gauge transformations to  general rank tensors charged under arbitrary number of gauge groups was given in {\bf II}.

From the field $\phi$ and the noncommutative Wilson lines one can construct a gauge invariant composite field
\begin{equation}
\Phi(x)=(W \otimes W) \phi(x).
\end{equation} 
With this gauge invariant object it is easy to construct  gauge invariant Lagrangians. A crucial 
observation is that $\Phi$ can be obtained from $\phi$ by a gauge transformation of the form
 (\ref{modified}) with $U(x)=W(x)$. 
 In other words $\phi$ and $\Phi$ lie in the same gauge orbit. Thus it seems that the gauge invariant Lagrangian 
 that is a function of $\phi$ and $W$ can be equally well considered as a function of a single field $\Phi$.
 This would correspond to fixing the gauge $U(x)=W(x)$, implying that any dependence on the Wilson lines
can be removed by mere gauge fixing.

Physical interpretation of the gauge transformations with Wilson lines inserted still remains somewhat obscure, 
since all dependence on the Wilson lines in the action seems to vanish after the gauge fixing.
Especially, it is not clear whether the gauge fixed field can be indeed treated as a single field without
any dependence on the gauge fields.
Reaching a complete understanding of this issue remains an open problem.

Another subtle point is the extra freedom in the construction of the Lagrangian that was pointed out in {\bf II}.
To construct the kinetic term e.g. for a scalar field transforming under the rank two tensor representation of $U_*(N)$,
one first defines the gauge invariant objects corresponding to the gauge field and covariant derivatives \cite{Chu:2001kq}
\begin{eqnarray}
\mathcal{A_\mu} &=& W*(A_\mu-i\partial_\mu)* W^{-1} \\
\mathcal{D_\mu} &=& (1\otimes 1)\partial_\mu + i(\mathcal{A_\mu} \otimes 1) +i(1 \otimes \mathcal{A_\mu}),
\label{}
\end{eqnarray}
where $A_\mu$ is the ordinary NC $U_*(N)$ gauge field that transforms as
\begin{eqnarray}
A_\mu \rightarrow  U* (A_\mu - i\partial_\mu) *U^{-1}.
\label{}
\end{eqnarray}
Then the kinetic part of the action can be defined as
\begin{eqnarray}
\int d^4 x \; \tr \left| \mathcal{D_\mu}*\Phi \right|^2.
\label{aktio}
\end{eqnarray}
It should be noted that as $\Phi$ is in fact gauge invariant, there is in principle
no reason to introduce gauge fields or covariant derivatives, as an ordinary derivative would do just as well. However,  (\ref{aktio}) has the property that it reduces to the ordinary gauge invariant kinetic term of the rank two $U(n)$ field $\phi$ in the commutative limit. If the ordinary derivative were chosen, there would be not trace of the gauge fields in the Lagrangian
after the gauge fixing.


\section{Model building}


\subsection{Model building based on the Seiberg-Witten map.}
Since the Seiberg-Witten map can be used to define noncommutative versions of any commutative gauge group, it can be used to construct
NC versions of particle physics models based on any commutative gauge field theory.
The Lagrangian is obtained as an expansion in $\theta^{\mu\nu}$
with the zeroth order corresponding to the commutative model and can be used to calculate NC corrections to physical processes.
In \cite{Calmet:2001na} a NC Standard Model based on the gauge algebra $u_{NC}(1)\times su_{NC}(2)\times su_{NC} (3)$
was constructed. The authors also argued that the charge quantization problem is avoided by this construction.
 In \cite{Brandt:2003fx} it was shown that models based on the Seiberg-Witten map are also anomaly-free.

New features in the model appear when the $\theta$-corrections are considered. For example, parity is broken and
vertices with coupling between the $SU(3)$ gauge bosons, $U(1)_Y$ gauge bosons and quarks appear.
Other new effects such as neutral decays of heavy particles such as $b$ and $t$ quarks may provide experimental signals
for NC space-time.
Since the Seiberg-Witten map allows also more general representations of the gauge algebra than those allowed by
the no-go theorem one can use this method to construct GUT models. In \cite{Aschieri:2002mc} a NC GUT based on the
gauge group $SU(5)$ was constructed.

Supersymmetric extension of the model has not been considered due to the difficulties in reconciling supersymmetry with the Seiberg-Witten map.
Alternative mappings from the NC SYM are obtained depending on whether one works in the superfield or component field formalism. The Seiberg-Witten map
for superfields suffers from nonlocal terms, while the ordinary Seiberg-Witten map for component fields realizes supersymmetry in the commutative side in a nonlinear form.
These issues have been recently reviewed and clarified in \cite{Martin:2008xa}.


\subsection{NC SM based on the no-go theorem}

If the noncommutative theory is taken to be fundamental, the gauge theory should be built
 and analyzed in terms of noncommutative concepts.
  The expansion of the action in  the Seiberg-Witten map approach may also lose important nonlocal effects arising from the noncommutativity that can
 be seen only if the complete *-products are present. To retain these effects it is desirable
 to work in the level of the noncommutative fieds without expanding the *-products. 

In \cite{Chaichian:2001py} a noncommutative version of the Standard Model of particle physics was constructed without resorting to the Seiberg-Witten map.
Even though the field content can be constructed based on $U_*(n)$ and in accordance of the no-go theorem, a novel feature is needed to eliminate extra degrees of freedom of the gauge fields and to obtain fractional charges for the quarks.
In this mechanism also the modified gauge transformations come into play and at least a small sector in the model has to circumvent the no-go theorem.

The starting point is the minimal noncommutative extension of the Standard Model's $SU(3)\times SU(2) \times U(1)$, i.e. $U_*(3)\times U_*(2)\times U_*(1)$. The gauge group of the NC theory has two extra degrees of freedom compared to the commutative counterpart, corresponding to the trace parts of $U_*(2)$ and $U_*(1)$. These extra degrees of freedom do not decouple properly at low energies and have to be suppressed somehow in order for the model to be in agreement with observations. In order to suppress the contribution of these extra degrees of freedom to low energy physics, a mechanism similar to the Higgs mechanism of electro-weak symmetry breaking can be applied. This consists of introducing extra scalar fields that couple only to the tr-U(1) parts of the $U_*(N)$ gauge groups and provide masses to the corresponding gauge fields through their nonzero vacuum expectation values. 
This mechanism provides also a way to overcome another obstacle posed by the no-go theorem, namely charge quantization.
After the symmetry reduction the  correct charge and field content is obtained. Thus the charge quantization problem turns out to be a virtue:
it explains why the charges of quarks are quantized to fractional values of the electric charge.
This can be also considered as an advantage over the model based on the Seiberg-Witten map.




For the symmetry reduction mechanism, a new field, the so-called Higgsac field, has to be introduced.
In \cite{Chaichian:2001py} the Higgsac field was defined as a scalar field that transforms only under the NC $U_n(1)$ part $U_*(n)$. 
The NC $U_n(1)$ is defined as $U_*(n)/$NC $SU(n)$, where NC $SU(n)$ is the part of $u_*(n)$ whose trace goes to zero at the limit of vanishing $\theta$ (details of the relevant definitions
can be found in \cite{Chaichian:2001py}).

The symmetry reduction in the NC Standard Model takes place in two stages. First, a Higgsac field $\Phi_1$ is used to reduce the  $U_*(3)\times U_*(1)$ part of the gauge group, resulting
in a mass term for a combination of the trace parts of the gauge bosons. A massless trace component also remains and is successively mixed with the trace part of the $U_*(2)$ gauge boson via coupling to another Higgsac field, $\Phi_2$. When $\Phi_2$ obtains a nonzero VEV, only one massless $U(1)$ gauge field remains.
With high enough values for the Higgsac VEVs, the massive tr-$U(1)$ fields can be decoupled from low-energy physics.

It was shown in \cite{Hewett:2001im} that the Higgsac mechanism as described in \cite{Chaichian:2001py} leads to violation of unitarity in scattering processes. As it was noticed in \cite{Chaichian:2004yw} and further elaborated in  {\bf I} and {\bf II}, this is due to the ill definition of the original Higgsac mechanism. The way that the $U_n(1)$ symmetry is defined and spontaneously broken does not respect the original noncommutative gauge symmetry of the theory and thus it is not spontaneous symmetry breaking.
A refined version of the Higgsac mechanism was developed in \cite{Chaichian:2004yw} and {\bf I}, {\bf II }. The refined construction resorts in the modified gauge transformations with NC Wilson lines.

Let us describe the refined Higgsac mechanism in more detail. 
Take for example the $U_*(n)$ gauge symmetry and consider a scalar field defined by:
\begin{eqnarray}
\Phi(x) = \frac{1}{n!}\epsilon_{i_1 i_2 ... i_n} W^{i_1}_{j_1} * W^{i_2}_{j_2} * ... * W^{i_n}_{j_n} * \phi^{j_1 j_2 ... j_n} (x).
\label{}
\end{eqnarray}
This is just the rank n-tensor field $\phi$ in the gauge $U=W$ and contracted with the antisymmtric tensor $\epsilon$.
The composite field is gauge invariant and can be expanded as
\begin{eqnarray}
\Phi(x)=\phi(x)+...,
\label{composite}
\end{eqnarray}
where the first term is the original Higgsac field
\begin{eqnarray}
\phi(x)=\frac{1}{n!}\epsilon_{i_1 i_2 ... i_n} \phi^{i_1 i_2 ... i_n}(x),
\label{}
\end{eqnarray}
that transforms only under the tr-$U(1)$ part of the gauge group and has charge $n$.
A potential providing a nonzero VEV for the composite field $\Phi$, will thus
break the tr-U(1) symmetry and provides a mass term for the corresponding gauge field.
Since the Lagrangian is invariant under the original $U_*(n)$ symmetry, unitarity
should be preserved when the gauge invariant completion in (\ref{composite}) is taken
into account.

In \cite{Chaichian:2004yw} also another problem with the NC Standard Model was addressed. With the matter content of the original model the triangle anomalies do not cancel. To remove this problem, two new leptonic $U_*(2)$ doublets were introduced to cancel the anomalies.
For these fields one can write Yukawa couplings to the Higgsac field that provides them masses.

The model described in this section leads to several novel features. Part of them arise from the *-products and vanish in the 
commutative limit $\theta\rightarrow 0$ but due to the group theoretical structure of the model there are also features that are not sensitive to the commutative limit. 
This is to be contrasted with the model based on the Seiberg-Witten map where all new effects are proportional to the noncommutativity parameter.
 For example, the two new massive gauge bosons provide NC corrections to the $\rho$-parameter that do not depend on $\theta^{\mu\nu}$.
Comparing these contributions with the usual commutative loop corrections one obtains a lower bound on the masses of the new gauge bosons 
\begin{eqnarray}
m_{G^0,W^0} \gtrsim 25 \, m_Z.
\label{}
\end{eqnarray}
A notable effect related to $\theta$ itself is the dipole moment proportional to $\theta$ that arises for all particles.
Experimental bounds on the neutrino dipole moment translate into bound on the scale of noncommutativity.
A crude estimation gives
\begin{eqnarray}
\Lambda_{NC} \gtrsim 10^3 \,GeV
\label{}
\end{eqnarray}
when compared with astrophysical experiments \cite{Chaichian:2001py}.
This is of the same order as the bounds arising from Lamb shift and bounds on Lorentz violation \cite{Chaichian:2000si, Carroll:2001ws}.

It should be noted that some subtle points in this construction still remain. First thing is to note that the matter fields
were originally constructed by exploiting the restrictions of the no-go theorem. However, with the Higgsac-construction described above,
the modified gauge transformations are needed in the model anyway and the no-go theorem has to be circumvented. Thus, in principle, there is no need to stick with
the ordinary gauge transformations for the matter sector either. Another unclear issue is the treatment of the modified gauge transformations themselves and especially the observation discussed in {\bf II}: it seems as if the composite Higgsac field can be considered as $\phi$ with a specific gauge fixing. Thus, without expanding the Wilson lines it seems that the scalar field does not couple to the gauge fields at all and
thus the symmetry reduction is in fact just an artifact of the expansion. Thus the issue of the symmetry reduction
should be considered as an unsolved problem of the model.

Another candidate for a noncommutative Standard Model was constructed in paper \cite{Khoze:2004zc}. The model is based on the 
larger gauge group $U_*(4)\times U_*(3)\times U_*(2)$. Without the supersymmetric enhancement discussed in paper {\bf I},
the $U_Y (1)$ gauge field of the model described above can not be treated as a photon due to disastrous quantum corrections. 
The idea behind the larger gauge group is to construct the $U_Y(1)$ gauge field from a traceless combination of $U_*(n)$ generators
and to make the tr-$U(1)$ parts decouple at low energies. It was argued in \cite{Khoze:2004zc} that this decoupling is actually produced by the UV/IR-mixing
which causes a logarithmic decreasing of the tr-U(1) coupling at low energies.
However, the logarithmic running of the couplings is too slow to provide a solution to the problem \cite{Jaeckel:2005wt}.

 

\subsection{NC MSSM}

The $U_*(3)\times U_*(2)\times U_*(1)$ -model described in previous subsection can be generalized to include supersymmetry. While supersymmetrization is interesting for the same reasons as in
the commutative case, it turns out that supersymmetry also improves some problems arising from noncommutativity at the quantum level.
Especially, quantum corrections to the polarization tensor of the  $U_Y(1)$ gauge boson include harmful effects 
such as vacuum birefringence and tachyonic instability \cite{Matusis:2000jf, AlvarezGaume:2003mb, Armoni:2001uw, Jaeckel:2005wt}. With the help of supersymmetry these problems can be alleviated.

The generalization of the superfield formalism to the noncommutative setting was given in \cite{Terashima:2000xq}. 
The noncommutative superspace is constructed by extending the noncommutative algebra of the bosonic coordinates to include the anticommuting fermionic superspace coordinates:
\begin{eqnarray}
 & [\hat{x}^m,\hat{x}^n]=i\theta^{mn}, & \nonumber  \\
 & [\hat{x}^m,\hat{\theta}^\mu]=0, & \label{SuperAlg}\\
 & \{\hat{\theta}^\mu,\hat{\theta }^\nu \} = \{\hat{\bar{\theta}}^{\dot{\mu}} , \hat{\bar{\theta}}^{\dot{\nu}}\} = \{\hat{\theta}^\mu,\hat{\bar{\theta}}^{\dot{\mu}}\} = 0,\nonumber 
\end{eqnarray}
where we have used the roman indices to denote the space-time directions.
Here we assume that the commutators of the fermionic coordinates are not deformed and thus there is no non-anticommmutativity. 
Then the superfields are defined just as in the commutative case and the algebra (\ref{SuperAlg}) can be realized by
imposing the Moyal *-product,
\begin{eqnarray}
 (fg)(\hat{x},\hat{\theta},\hat{\bar{\theta}})\longmapsto (f*g)(x,\theta,\bar{\theta})=e^{{i\over
  2}\Theta^{mn}\partial_{x_m}\partial_{y_n}}
 f(x,\theta,\bar{\theta})g(y,\theta,\bar{\theta}){\Bigg |}_{x=y}.
\label{}
\end{eqnarray}
The generalization of supersymmetric gauge theories to the noncommutative setting goes through as 
in the non-supersymmetric case and the restrictions of the no-go theorem apply.
Construction of the noncommutative superpotential for MSSM leads to two possible choices
that give the commutative MSSM matter content with some additional fields included. One of the choices
include two leptonic doublets for each family to cancel chiral anomalies, while the other includes four such doublets. In Table 4.1 the matter content for the minimal model with two leptonic doublets $L'_i$ and $L''_i$ is given, and the corresponding superpotential is
\begin{eqnarray}
{\cal W}&=&\lambda_e^{ij}H_1 * L_i * E_j+\lambda_u^{ij}Q_i * H_2 * \bar{U}_j
 +\lambda_d^{ij}Q_i * H_3* \bar{D}_j\nonumber\\
 &+&\mu_{12} H_1 * H_2+\mu_{34} H_3 * H_4 \nonumber \\
&+&\left( \alpha_1^{ijk}Q_i*L_j*\bar{D}_k +\alpha_2^iL_i*H_4 +\alpha_3^{i}L_i^\prime*H_1 \right. \label{yukawa}\\
 &+&\left. \alpha_4^iL_i^{''}*H_4+\lambda_{L^{\prime\prime}}^{ij} H_1 * L^{\prime\prime}_i * E_j \right)\,.\nonumber
\end{eqnarray}
The model has $U_*(3)\times U_*(2)\times U_*(1)$ symmetry which has to be reduced.
The symmetry reduction mechanism with Higgsac fields has to be generalized to the sypersymmetric setting and  to this aim one has to first introduce
supersymmetric noncommutative Wilson lines. The supersymmetric generalization of the Higgsac mechanism turns out to be rather straightforward and without going to the details, which can be found in paper {\bf I}, the $R$-parity
conserving superpotential reads
\begin{eqnarray}
 &&{\cal W}={\cal W}_{\mbox{\tiny Yukawa}}+{\cal W}_{\mbox{\tiny Higgsac}}\,, \\
 &&{\cal W}_{\mbox{\tiny Yukawa}}=\lambda_e^{ij}H_1 * L_i * E_j
   +\lambda_u^{ij}Q_i * H_2 * \bar{U}_j+\lambda_d^{ij}Q_i * H_3* \bar{D}_j
\nonumber \\
&&~~~~~~~~~~~~+\mu_{12} H_1 * H_2+\mu_{34} H_3 * H_4
+\lambda_{L^{\prime\prime}}^{ij} H_1 * L^{\prime\prime}_i *E_j\,,\\
&&{\cal W}_{\mbox{\tiny Higgsac}}=\sum_{a}\left(m^2\Phi_a
 -\frac{\lambda}{3}\Phi_a*\Phi_a*\Phi_a\right)\,,
\end{eqnarray}
where the index $a$ denotes groups the Higgsac superfields are associated with,
 $a=U_*(2)\times U_*(1), U_*(3)\times U_*(2)\times U_*(1)$ .
\begin{table}
\begin{center}
\begin{tabular}{cccc}
\hline
Chiral Superfield & $U_\star(3)$ & $U_\star(2)$ & $U_\star(1)$ \\
\hline
$L_i$ & 1 & 2 & 0\\
$\bar{E}_i$ & 1 & 1 & $-1$\\
$Q_i$ & 3 & $\bar{2}$ & 0\\
$\bar{U}_i$ & $\bar{3}$ & 1 & +1\\
$\bar{D}_i$ & $\bar{3}$ & 1 & 0\\
\hline
$L_i^\prime$   & 1 & 2 & $-1$\\
$L_i^{\prime\prime} $ & 1 & 2 & 0\\
\hline
$H_1$ & 1 & $\bar{2}$ & +1\\
$H_2$ & 1 & 2 & $-1$\\
$H_3$ & 1 & 2 & 0\\
$H_4$ & 1 & $\bar{2}$ & 0\\
\hline\\
\end{tabular}
\caption{Matter content of MSSM. The index $i$ denotes the family.}
\end{center}
\label{MSSMmatter}
\end{table}
As in the NC SM, here also the model possesses new features compared to ordinary MSSM, that are not related to the scale of noncommutativity. Instead of two Higgs doublets, four had
to be introduced to build the superpotential. In this model the new down-type Higgs bosons $H_3$ and $H_4$ must obtain vacuum expectation values in order to give masses to the down-type quarks, while $H_1$ and $H_2$, corresponding to the usual Higgs bosons appearing in the commutative MSSM, provide masses to the up-type quarks and leptons.
The new Higgs fields can provide indirect observable signals of noncommutativity.

An important feature of supersymmetry is that the dangerous  quantum corrections to the polarization tensor of the tr-U(1) field cancel. Thus the infrared singularity, tachyonic mass and vacuum
birefringence problem are removed. However, SUSY has to be broken at the energy scales of the Standard Model, and contributions to these problems potentially arise from the scales with broken SUSY.
In paper {\bf I} the quantum corrections to the polarization tensor of the tr-$U(1)$ gauge boson in the presence of soft SUSY breaking terms were analyzed. When the bosonic and fermionic degrees of freedom match, the infrared singularity is cancelled and thus soft SUSY breaking does not affect this cancellation \cite{Matusis:2000jf}.
Other effects remain, but a qualitative analysis leads to the result that even with very conservative bounds on the scale of noncommutativity and the scale of supersymmetry breaking, the vacuum birefringence effect
can be highly suppressed. 

The dispersion relation for the polarization that receives the new effect can be written as
\begin{eqnarray}
 \omega^2-c^2\left({1 \over 1+\Delta n}\right)^2(k^3)^2=0,
\label{}
\end{eqnarray}
where 
\begin{eqnarray}
\Delta n&\simeq& {1 \over 8 \pi^2 \Pi_1}{\Delta M_{SUSY}^2 M_{SUSY}^2
  \over M_{NC}^4}\,.
\label{}
\end{eqnarray}
Here $M_{NC}$ is the scale of noncommutativity, $M_{SUSY}$ is the scale where SUSY breaking occurs and $\Delta M_{SUSY}^2$ describes the difference between bosonic and fermionic masses
\begin{eqnarray}
\Delta M_{SUSY}^2= \Delta M_{SUSY}^2={1\over 2}\sum_s
  M_s^2-\sum_f M_f^2. 
\label{}
\end{eqnarray}
The strongest bound on $\Delta n$ come from cosmological observations
\begin{eqnarray}
 |\Delta n_{\mbox{\tiny cosmo}}|\le 10^{-37}-10^{-32}\,.
\label{}
\end{eqnarray}
For example, choosing $M_{SUSY} \sim 10^{10}$, $M_{NC} \sim 10^{18}$, $m_j \sim 10^2$ and $k \sim 100$ GeV leads to
\begin{eqnarray}
 \Delta n \sim 10^{-62}\, ,
\end{eqnarray}
which is well beneath the experimental bounds.
Thus we conclude that the existence of supersymmetry at high energies can suppress significantly the Lorentz violating
effects that seem to ruin the nonsupersymmetric theory. The behavior of the $U(1)$ running coupling constant is still altered by
UV/IR mixing and a more thorough analyzis of this behaviour is needed in order to derive phenomenological implications.
It should be also noted that this analysis does not remove the possibility of a tachyonic photon.
Finally, we note that the subtleties with the symmetry reduction mechanism described in the end of the previous sections remain also in the supersymmetric version.

Further extension of this model can be considered. The general representations made possible by the modified gauge transformations allow for construction of  noncommutative Grand Unified Theories. 
A NC GUT based on the $U_*(5)$ was briefly analyzed in \cite{Chu:2001kq}.

%% file: conclusions.tex
\chapter{Conclusions}

The necessity of modification of the description of space-time as a manifold at very short distances
calls for candidates for a theory of quantum space-time.
The idea of noncommutativity as a way to describe properties of the quantized space-time stems from the principles of quantum mechanics and gravity theory applied on physics
at very short distances \cite{Doplicher:1994zv}. Further support to this idea comes from open string theory in the presence
of a background field \cite{Seiberg:1999vs}. While the complete theory of physics at short distances is presumably
more complicated, quantum field theory in noncommutative space-time may capture some essential
features of the quantum space-time and provide insight into quantum gravity.

The papers included in this thesis encompass various aspects of quantum field theory and gauge symmetries in noncommutative space-time, enlightening
also some of the fundamental problems that still need solving.
One of the most important issues is the symmetry of quantum space-time. The canonical approach, with Heisenberg-like commutation
relations for the coordinates with a constant parameter of noncommutativity seem to break Lorentz invariance of the theory.
Then the Poincar\'e symmetry is preserved only in a twisted form \cite{Chaichian:2006we}.
Lorentz symmetry has become such a fundamental property in modern particle physics theories, that giving
it up seems awkward and thus attempts towards a formulation that preserves the Lorentz symmetry have been considered
in the literature \cite{Carlson:2002wj}. To preserve Lorentz invariance, it seems necessary to allow states with noncommuting space and time.
 On the other hand, time-space noncommutativity is known to lead to difficulties in  quantum field theory, as causality
 and unitarity is lost. Thus the Lorentz-invariant NC QFT also suffers from these problems {\bf III}.

 Another important flaw in the formulation of quantum space-time in terms of the canonical commutator relation is that
 the commutator induces infinite nonlocality. Infinite nonlocality causes effects in field theory that are difficult to reconcile
 with observed physics. 
The causality condition is modified to allow infinite propagation speed in at least one diretion. Also quantum corrections in field theory
suffer from a mixing between ultraviolet and infrared degrees of freedom that make renormalization difficult.
First attempt towards a formulation with restricted nonlocality can be found in \cite{Bahns:2006cp}.

 Despite the fact that some fundamental issues in NC theories are still to be solved, it is important
 to make contact with physics at accessible energies. If Lorentz invariance is indeed broken
 in noncommutative space-time, then it has important phenomenological consequences, which may provide
 a way to detect noncommutativity. Of utmost importance to model building is also to find the proper NC generalization of gauge symmetries.
 A straightforward generalization of the gauge transformations to include the *-products causes several restrictions on model building.
 These restrictions can be advantageous, providing an explanation for the charge quantization, but they also pose significant
 problems for the NC Standard Model \cite{Chaichian:2001py} and NC MSSM {\bf I}. In order to obtain the $SU(n)$ gauge symmetries of the Standard Model at low energies, it seems that at least part of the restrictions have to be circumvented.
One solution is to insert appropriate Wilson lines into the gauge transformations as proposed in \cite{Chu:2001kq} and {\bf II}. 
This approach allows one to circumvent some of the restrictions posed on the representations of noncommutative gauge groups.
Another popular approach to NC model building is to use the Seiberg-Witten map to define the NC gauge theory in terms of a commutative one. 

The study of noncommutative field theories provides fresh insight to physics at the fundamental level where the usual concept of particles
as fields in a continuous space-time can not be maintained. While some naive expectations have proven false, noncommutativity has certainly
brought new exciting features to the study of quantum field theories. We have found that fundamental properties of ordinary quantum field theories,
such as Lorentz symmetry and causality, need revision in noncommutative space-time. Finding a firm understanding of such issues will be important  
when paving the way towards a complete theory of quantum space-time.



